\documentclass[]{achemso}

\usepackage{lipsum} 
\usepackage[dvipsnames]{xcolor} 
\usepackage{graphicx} 
\usepackage{amsmath, amssymb} 
\usepackage{bm} 
\usepackage{xr} 
\usepackage{makecell}
\usepackage{changepage}

\usepackage{nameref}
\usepackage{varioref}
\usepackage{hyperref}
\usepackage{cleveref}

\definecolor{darkorange}{HTML}{D45500}

\title{Central role of filler--polymer interplay in nonlinear reinforcement of elastomeric nanocomposites}
\author{Pierre Kawak}
\affiliation{Department of Chemical, Biological, and Materials Engineering, University of South Florida, Tampa, Florida 33612}
\author{Harshad Bhapkar}
\affiliation{Department of Chemical, Biological, and Materials Engineering, University of South Florida, Tampa, Florida 33612 }
\author{David S. Simmons}
\affiliation{Department of Chemical, Biological, and Materials Engineering, University of South Florida, Tampa, Florida \ 33612  }
\date{\today}
\email{dssimmons@usf.edu}

\newcommand*{\abstracttext}{
Nanoparticles can greatly enhance the mechanical response of elastomeric polymers essential to a wide range of applications, yet their precise molecular mechanisms of high-strain reinforcement remain largely unresolved.
In particular, longstanding questions endure over the extent to which glassy bridges or tie chains between particles are needed to facilitate reinforcement.
Here we show, based on molecular dynamics simulations, that high-strain reinforcement emerges from an interplay between granular nanoparticulate compressive behavior in the normal direction and polymer incompressibility.
This feedback loop, which is initiated by a mismatch in the Poisson ratios of nanofiller and polymer, invokes a contribution from the polymer's bulk modulus to the elongational stress, while the tendency of the polymer to contract in the normal direction maintains a near-jammed filler state. This effect persists even once the direct filler elongational contribution becomes dissipative after the ‘Payne effect’ yield, as a consequence of enduring filler normal stresses mediated by direct particle-particle contacts.
These results indicate that direct particle-particle contact effects, even in the absence of potential augmenting mechanisms such as glassy polymer bridges, can drive the mechanical reinforcement effects typical of experimental systems.
}


\makeatletter
\renewcommand{\acs@tocentry@print}[1]{%
  \gdef\acs@tocentry@text{\small{#1}}%
  \AtEndDocument{%
  \if@twocolumn
      \@restonecoltrue\onecolumn
    \else
      \@restonecolfalse\newpage
    \fi
    \if@restonecol
      \twocolumn
    \else
      \newpage
    \fi
  }%
}

\renewcommand*{\acs@maketitle@extras}{%
  \acs@maketitle@extras@hook
  \acs@tocentry@print@aux
  \newpage
}
\makeatother
\let\oldmaketitle\maketitle
\let\maketitle\relax

\begin{document}

\begin{tocentry}
  \includegraphics[]{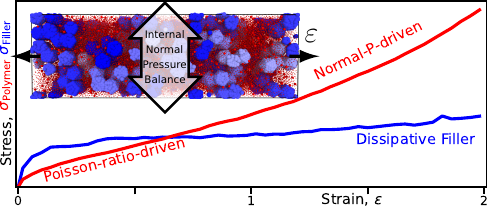}
  \\This study elucidates the molecular mechanisms of enhanced durability of elastomeric nanocomposites when subjected to large deformations.
Simulations reveal that a feedback loop between the elastomer matrix and nanoparticle network is the primary driver of reinforcement.
This reinforcement persists beyond traditional limits of filler yield, i.e., the Payne effect, suggesting a novel mechanism for material reinforcement.
\end{tocentry}

\begin{@twocolumnfalse}
  \oldmaketitle

  \begin{abstract}
    \abstracttext
  \end{abstract}
\end{@twocolumnfalse}

\section{Introduction}

Elastomeric nanocomposites are of wide interest due to their capacity to exhibit enhanced properties relative to neat elastomers.~\cite{Hamed2000, Song2016, Quaresimin2016, Kumar2017, Kumar2017a}
They are employed in a wide range of settings requiring robust mechanical response under mechanical load~\cite{Roland2016} and can also introduce additional orthogonal functionalities such as modified thermal or electrical conductivity.~\cite{Kumar2017, Kumar2017a}
Elastomeric nanocomposites are arguably the oldest synthetic nanocomposites, beginning at large scale in auto and aero tires comprised of vulcanized rubber interspersed with carbon black nanoparticles (NPs), and subsequently finding applications in coatings, seals, dampers, cushions, and transport belts.~\cite{Wang2005}
In those contexts, the addition of carbon black dramatically reinforces the elastomeric matrix.

This reinforcement involves a cascade of effects with increasing strain.~\cite{Robertson2021} In the linear regime, nanofillers can increase the composite's Young's modulus by an order of magnitude or more relative to neat elastomers. Filled elastomers then exhibit a soft yield event at \textit{ca.} 10-20\% strain, known as the Payne effect,~\cite{Fletcher1954, Payne1962} which is absent in neat elastomers and signals the onset of large additional dissipative contributions to the stress.
Finally, NPs strongly modify high-strain strain-hardening behavior as well as the void nucleation and growth events that precede failure.~\cite{Steck2023} Collectively, these effects can enhance fracture strength from \textit{ca.} 2 MPa to \textit{ca.} 25 MPa or more.~\cite{MEDALIA1994, Robertson2021}
Despite their immense economic impact, and their potential in new transformative materials development, the precise mechanisms by which introduction of NPs yields immense mechanical reinforcement in elastomers remain unsettled.~\cite{Song2016, Kumar2017}

Historically, reinforcement in composites was understood based on Einstein-like models for reinforcement in the dilute regime, and series expansions of this approach to semi-dilution.~\cite{Einstein1905, Guth1945, Mullins1965, Pryamitsyn2006}
However, this approach dramatically underpredicts the extent of reinforcement at the loadings relevant to most applications.~\cite{Smith2017}
More recent work has suggested that a central reason for this failure is a fundamentally non-dilute behavior of reinforcing particles: reinforcement is dominated by formation of a percolated or jammed filler network.~\cite{Boschan2016, Robertson2021, Huang2022, Steck2023}
However, this raises a new question.
Cohesive interactions between the NPs are expected to be relatively weak in comparison to the elastomer, given that the elastomer is interconnected with covalent bonds, whereas particles interact via relatively weak van der Waals interactions. How can these weak interparticle cohesive interactions yield appreciable reinforcement during large extensional strain? A useful analogy here is concrete, which consists of a largely non-cohesive filler network bound together with a weak binder matrix. Concrete is extremely strong under compression but notoriously weak under tension because tension dejams the particle network. Why, then, do elastomers with at most weakly cohesive nanofiller exhibit strong reinforcement under high extensional strain?

A dominant perspective in the field is that NPs in elastomeric matrices exhibit an emergent cohesive interaction as a consequence of the formation of glassy bridges of polymer between the particles.~\cite{Dannenberg1986, Berriot2003, Merabia2008, Papon2012, Starr2016, Sotta2017, Tauban2017, Huang2022}
The idea is that these glassy bridges essentially glue together the particle matrix, dramatically enhancing the reinforcement and the response under extensional strain.
Indeed, it is now well established that strongly attractive rigid interfaces can produce substantial enhancements in the glass transition temperature $T_g$ of proximate polymer over length scales on the order of 10 nanometers.~\cite{DiazVela2020} Recent experimental evidence from poly(methyl methacrylate)--silica nanocomposites suggests that these interfacial dynamical gradients can correlate with mechanical reinforcement in at least some cases.~\cite{Ondreas2019}
However, there is considerable debate over whether this effect is relevant under the conditions associated with nanocomposite elastomer deformation ($T >> T_{g}$).
Furthermore, because these effects vanish or even reverse for weak polymer--filler interactions, at a minimum there must be some elastomeric nanocomposites -- those that are relatively weakly interacting -- in which these effects are absent.

An alternate earlier scenario suggests that reinforcement is primarily mediated through direct filler--filler contacts rather than via glassy bridges.
In 1993, Witten, Rubinstein, and Colby (WRC) proposed a theoretical picture wherein this direct contact model could still lead to an enormous mechanical reinforcement under large elongational strain.~\cite{Witten1993}
While the WRC model treats the mechanics of a percolating rather than jammed NP network, we focus here on their central proposition.
Essentially, the incompressibility of the elastomeric matrix is proposed to dominate the boundary conditions of deformation in the normal direction.
As a consequence, elongation of the composite is necessarily accompanied by compression of the filament matrix in the normal direction and it is this normal compression that is postulated to be the primary driver of reinforcement under high strain.

Our prior work provided additional insight towards a mechanism along these lines with an insight in the linear regime~\cite{Smith2017, Smith2019}: the key driver of linear elongational reinforcement is not precisely elastomer \emph{incompressibility} but rather a preferred Poisson ratio \emph{mismatch} between the filler network and elastomer matrix. Unlike in concrete, the elastomeric matrix and jammed nanoparticulate medium prefer to deform under quite distinct Poisson ratios - approximately 0.5 for the elastomeric matrix, but considerably less for the co-continuous nanoparticulate medium.
The lower Poisson ratio preferred by the filler results from the locally rigid nature of the particle, combined with their percolation or jamming to form a particulate network over larger scales.
The resulting `compromise' Poisson ratio is intermediate between these two values, which forces the elastomeric matrix to undergo an appreciable increase in volume with elongation.
This volume increase invokes an appreciable contribution from the elastomer's bulk modulus, which is roughly 1000 times its neat Young's modulus,~\cite{Holownia1975} to the composite's Young's modulus.~\cite{Smith2019}
This finding was also consistent with experimental work reporting on Poisson ratio shifts in filled elastomers~\cite{Robertson2007} and even nanofilled oils.~\cite{Warasitthinon2018}
Moreover, while suggesting a need for corrections to the WRC assumption of perfect incompressibility during deformation and some of its implications, this scenario preserves the WRC's central proposed role for lateral compression of the filler particles;
it is the outward normal pressure imposed by the filler network under elongation that leads to elastomer matrix volume deviations and thus leads to elongational contributions from its bulk modulus.

While these findings clarify the leading-order mechanism of low-strain elastomer reinforcement by nanoparticles, they leave open the question of how this mechanism can extend to higher strains beyond the yield of the initial filler network.
Indeed, while the  WRC theory suggests that the filler network remains jammed/percolated due to lateral compression during uniaxial extension, it did not yet account for filler dejamming or denetworking events that accompany (and likely drive) the Payne effect.~\cite{Wang2005, Robertson2005, Robertson2006, Robertson2007, Richter2010, Liu2010}

To understand the mechanism of reinforcement in this high strain regime in the absence of strong filler cohesive interactions, here we report on molecular dynamics (MD) simulations of nonlinear deformation of a nanofilled elastomer in the unentangled regime using the LAMMPS software.~\cite{Intveld2008, Brown2011, Thompson2022}
We focus on a loading (26.1\% filler by volume) that is well beyond the dilute regime but relatively early in the regime of filler network formation and strong reinforcement.~\cite{Baeza2016}
This enables us to probe the physics of reinforcement at a point at which the stress response of the filler network is not so large as to swamp all other contributions, providing a window on multiple mechanisms that interact to produce the overall response.
We compare this system to simulations of the same crosslinked elastomer in the neat state (without any NPs).
Crucially, we simulate systems in which particle--polymer pairwise interactions are symmetric with polymer--polymer interactions.
Combined with the choice of a temperature far above $T_g$ ($T = 1 >2.5 T_g=$ in Lennard-Jones, LJ, units),~\cite{Hung2019} this avoids glassy bridge effects that can emerge at lower temperatures.~\cite{Fakhraai2008, Daley2012, DiazVela2018, DiazVela2020, Schweizer2019}
Our results here are thus specific to elastomers lacking such effects -- a scenario expected to commonly emerge in systems involving low-$T_g$ elastomers and lacking especially strong polymer--particle attractive interactions.

Moreover, our simulations are performed using a thermostat that tends to damp hydrodynamic effects,\cite{Smith2017, Smith2019} which are known to be second order~\cite{Robertson2021} as described above but that could otherwise serve as a confounding factor in studying jamming-related phenomena.
Consistent with the physics discussed above, a constant zero normal stress is maintained during deformation to allow the system to conform to its preferred Poisson ratio, $\nu$.
More details of the simulation methods, including annealing procedure and deformation rates,~\cite{Martinez2009} can be found in the Methods Section and in prior works where these methods were established.~\cite{Smith2017, Smith2019}

\section{Methods} \label{sec:methods}


To study filled elastomers, we utilize uniaxial extension molecular dynamics (MD) simulations using the open-source Large-scale Atomic/Molecular Massively Parallel Simulator (LAMMPS) software.~\cite{Intveld2008, Brown2011, Thompson2022}
We analyze the resulting stress response upon uniaxial extension at a constant strain rate of $5\times10^{-5}/\tau_{LJ}$ using a MD timestep of $10^{-3} \tau_{LJ}$, where $\tau_{LJ}$ is the LJ unit of time, approximately equal to a picosecond in real units.
This rate was chosen based on prior study as within a range of rate-independence of the stress response for this model polymer.~\cite{Smith2017, Smith2019}
We resolve the stress response on a per-species basis, as well as spatially relative to filler particles, to further understand the origins of reinforcement in nanofilled elastomers.

We stretch and compare two systems.
First, we study a neat crosslinked polymer made up of 5000 end-crosslinked chains with 20 Kremer-Grest (KG) beads at 95\% extent of crosslinking.~\cite{Stevens2001, Tsige2004}
Second, a filled polymer matrix constituted by the same amount of polymer in addition to filler clusters, each comprised of 7 icosahedral particles sintered into a random cluster configuration (see below) at a 26.1\% filler by volume (75 pounds per hundred rubber).


\subsection{Initial Configuration Preparation and Annealing Procedure} \label{sec:init}


\subsubsection{Polymer Model}

Our simulations feature 5000 polymer chains with 20 coarse-grained KG beads.
All beads in our simulations interact via the LJ nonbonded potential,
\begin{equation} \label{eq:LJ}
E_{LJ}(r) = \begin{cases}
4 \epsilon \left[ \left( \frac{\sigma}{r} \right)^{12} - \left( \frac{\sigma}{r} \right)^{6} \right] & r \le r_{c} \\
0 & r > r_{c},
\end{cases}
\end{equation}
where $\epsilon$ and $\sigma$ are the energy and length scale used in our simulations, and $r_{c}=2.5\sigma$ is the cutoff radius for polymer--polymer and polymer--filler interactions.
Polymer bonds are modeled with a finite extensible nonlinear elastic (FENE) bond potential,
\begin{equation}
E_{\mathrm{FENE}} = -0.5 \kappa R_0^2 \ln \left[1-\left(\frac{l}{R_0}\right)^2\right] + E_{LJ}\left(l\right) + \epsilon,
\end{equation}
where $l$ is the distance between two consecutively bonded LJ beads, $\kappa = 30\epsilon/\sigma^2$ is the strength of the FENE spring, $R_{0} = 1.5\sigma$ is the maximum FENE bond length, and $E_{LJ}$ is computed using Eq.~\ref{eq:LJ} with a cutoff of $r_{c}=2^{1/6}\sigma$.

\begin{table}
\centering
\begin{tabular}{ |p{3.5cm}|p{1cm}|p{1cm}|p{1cm}| }
\hline
Interaction pair & $\epsilon$ & $\sigma$ & $r_{c}$ \\
\hline
Polymer--Polymer & $1.0$ & $1.0$ & $2.5$\\
Polymer--Filler & $1.0$ &  $1.0$ & $2.5$\\
Filler--Filler & $1.0$ & $1.0$ & $2^{1/6}$\\
\hline
\end{tabular}
\caption{Pair potential parameters in filled elastomer simulations.}
\end{table}

These polymer chains are end-crosslinked via a standard model used by our prior work and others.~\cite{Duering1994, Svaneborg2005}
Briefly, 2500 crosslinking KG beads are introduced to the polymer melt, which form bonds with chain end beads when they are within $1.3\sigma$ distance.
Such distance checking occurs every ten MD steps with LJ timestep $\delta t = 0.001 \tau_{LJ}$.
Every crosslinker bead can bond to 4 end beads, while each end bead can only bond with one crosslinker.
With $2500$ crosslinker beads, $10000$ potential crosslinking bonds can be formed.
The simulation is stopped after 95\% crosslinking is achieved, yielding a well-developed crosslinked network.~\cite{Smith2017}

\subsubsection{Model Filler Particles}

In this study, we use fractal clusters of seven icosahedrally-shaped nanoparticles to model nanofiller within our polymer matrix, in accord with previous work in our group.~\cite{Smith2017, Smith2019}
An individual icosahedral particle constitutes a center bead surrounded by three icosahedral shells of beads, making up $147$ total beads with $936$ harmonic bonds.
Filler beads interact with each other via the purely-repulsive LJ potential, i.e., with Eq.~\ref{eq:LJ} with $r_{c}=2^{1/6}\sigma$.
All bonds between filler beads are modeled via a harmonic potential, as shown by Eq.~\ref{eq:harm_bond},
\begin{equation} \label{eq:harm_bond}
E_{\mathrm{harm}} = \kappa_{\mathrm{harm}} \left( l - l_{0} \right)^{2},
\end{equation}
where $\kappa_{\mathrm{harm}}=1000\epsilon/\sigma^{2}$ is the bond stiffness and $l_{0}$ is the equilibrium bond length.
Within the innermost shell surrounding the center bead, beads are bonded to each other with $l_{0}=\sigma$ and to the central bead with $l_{0} = 0.951\sigma$.
Similarly, each shell’s beads are bonded with surrounding beads of the same shell ($l_{0}=\sigma$), as well as bonded with nearest beads within neighboring shells and second-closest beads within neighboring shells with $l_{0} = 0.9510\sigma$ and $1.380\sigma$, respectively.
Filler bond parameterization is summarized in Table~\ref{tbl:harm_bond}.

\begin{table}
\centering
\begin{tabular}{ |p{2.85cm}|p{1.70cm}|p{1.70cm}|p{1.70cm}| }
\hline
Interaction pair & $\kappa$ $[\epsilon/\sigma^{2}]$ & $l_{0}^{\mathrm{nearest}}$ $[\sigma]$ & $l_{0}^{\mathrm{cross}}$ $[\sigma]$ \\
\hline
Center -- Inner & $10^{3}$ & 0.951057 & N/A \\
Inner -- Middle & $10^{3}$ & 0.951057 & 1.380039 \\
Middle -- Outer & $10^{3}$ & 0.951057 & 1.380039 \\
Sinters         & $10^{3}$ & 1.000000 & 1.414214 \\
Same Shell      & $10^{3}$ & 1.000000 & N/A \\
\hline
\end{tabular}
\caption{Harmonic bond potential parameters of sintered filler particles in filled elastomer simulations.}
\label{tbl:harm_bond}
\end{table}

A sintering algorithm, previously developed by our group, is used, wherein a random face of an icosahedron is chosen and another icosahedron is added to this face and bonded ($l_{0} = 1.4142\sigma$).~\cite{Smith2017}
We refer to the number of icosahedral primary particles in a sintered cluster via the notation Np$k$, where $k$ is the number of sintered particles. For example, Np$7$ refers to a cluster containing 7 sintered icosahedral particles.
The algorithm that randomly generates these clusters was run to create a large number of unique clusters ($10^{4}$) and the length to diameter of the particle was determined.
Using Gaussian distribution, the quartile with highest length-to-diameter ratio (elongated as opposed to spherical) were not used in order to exclude non-physically elongated clusters that might unrealistically dominate stress response.
A single icosahedral particle maps to a diameter in the $7-14$ nanometer range while the longest length scale of the Np7 particle is in the $15-45$ nanometer range.~\cite{Smith2017}

\subsubsection{Dispersion of Filler Particle Clusters within Elastomer Matrix} \label{sec:init:anneal}

We model highly dispersed filler configurations via a previously established simulation protocol, which is summarized in Table~\ref{tbl:protocol}.~\cite{Smith2017}
First, the system is populated with the desired fraction of filler particles, along with polymer chains and crosslinker beads in a large box using Packmol.~\cite{Martinez2009}
Initially, the interactions between filler particles and polymer are turned off and polymer--polymer and filler--filler interactions are purely repulsive (Eq.~\ref{eq:LJ} with $r_c = 2^{1/6}\sigma$).
The filler anneals with a repulsive potential under these condition at high temperatures ($T=100$ in LJ units) to uniformly fill the box boundaries ($500\tau_{LJ}$) while the polymer anneals at a lower temperature ($T=1$ in LJ units).
The system boundaries are then shrunk to the estimated combined volume of neat polymer and the displaced polymer volume by the filler clusters over $2500\tau_{LJ}$.
The displaced polymer volume by filler is determined separately by calculating the volume displaced by each single cluster in the polymer system compared to the neat polymer volume.
Subsequently, the filled system is annealed to make a uniform distribution of filler and polymer for $2000\tau_{LJ}$ in the NVT ensemble.
A soft repulsive cosine potential between the filler particle centers is then introduced, with a prefactor equal to $5\epsilon$ and a cutoff distance of $3.15\sigma$, and the system is then annealed for $2000\tau_{LJ}$ (step 4).

\begin{table*}
\centering
\begin{tabular}{ |p{0.2cm}|p{5.5cm}|p{1.90cm}|p{1.90cm}|p{1.90cm}|p{0.9cm}|p{0.9cm}|p{0.9cm}| }
\hline
     & Ensemble                                            & Filler--Filler        & Polymer--Filler              & Polymer--Polymer      & $T_{F}$ $[\epsilon/k_{B}]$ & $T_{P}$ $[\epsilon/k_{B}]$ & $\tau$ $[\tau_{LJ}]$  \\\hline
1    & NVT polymer; NVE rigid filler                       & $r_{c}=2^{1/6}\sigma$ & -                            & $r_{c}=2^{1/6}\sigma$ & $100$                      & $1$                        & 500                   \\\hline
2    & NVT polymer; NVE rigid filler, $\rho \to \rho_0$    & $r_{c}=2^{1/6}\sigma$ & -                            & $r_{c}=2^{1/6}\sigma$ & $100$                      & $1$                        & 2500                  \\\hline
3    & NVT polymer; NVE rigid filler                       & $r_{c}=2^{1/6}\sigma$ & -                            & $r_{c}=2^{1/6}\sigma$ & $100$                      & $1$                        & 2000                  \\\hline
4    & NVT polymer; NVE rigid filler                       & soft                  & -                            & $r_{c}=2^{1/6}\sigma$ & $100$                      & $1$                        & 2000                  \\\hline
5    & NVE                                                 & $r_{c}=2^{1/6}\sigma$ & soft; $r_{c}=2^{1/6}\sigma$  & $r_{c}=2^{1/6}\sigma$ & $0$                        & $1$                        & 5000                  \\\hline
6    & \makecell[l]{NPT polymer;\\NVE rigid dilate filler} & $r_{c}=2^{1/6}\sigma$ & $r_{c}=2^{1/6}\sigma$        & $r_{c}=2.5\sigma$     & $1$                        & $1$                        & 600                   \\\hline
7    & NPT; crosslinking                                   & $r_{c}=2^{1/6}\sigma$ & $r_{c}=2^{1/6}\sigma$        & $r_{c}=2.5\sigma$     & $0$                        & $1$                        & $\sim 10^{4}$         \\\hline
8    & NPT                                                 & $r_{c}=2^{1/6}\sigma$ & $r_{c}=2.5\sigma$            & $r_{c}=2.5\sigma$     & $1$                        & $1$                        & 5000                  \\\hline
9    & NVT                                                 & $r_{c}=2^{1/6}\sigma$ & $r_{c}=2.5\sigma$            & $r_{c}=2.5\sigma$     & $1$                        & $1$                        & 5000                  \\\hline
\end{tabular}
\caption{Steps of filler dispersion into polymer matrix.}
\label{tbl:protocol}
\end{table*}

The filler particles are then frozen in place while the repulsive polymer beads continue to anneal.~\cite{Smith2017}
Because the simulations up to this point do not include polymer--filler interactions of any kind, the polymer and filler beads occupy the same space.
We thus next gradually grow the filler particles into the polymer matrix by pushing the polymer beads out from the volume occupied by the fillers.
To do so, a repulsive interaction between the filler's center bead and the polymer is introduced using a soft potential with $150\epsilon$ energy scale.
This soft potential's effective size is gradually expanded to a maximum of $3.8\sigma$, i.e., until the entire volume of each icosahedral particle is enveloped by its central bead.
To avoid non-physical rapid bead movement due to large-energy overlaps, the polymer's beads are restricted to small maximum movements every time step.
All the polymer beads contained within the filler particles are expelled in this way over a period of $4000\tau_{LJ}$ in the NVE ensemble.~\cite{Smith2017}
Then, interactions between the inner and outer shell filler beads and polymer beads are turned on gradually over $500\tau_{LJ}$ from $0$ to $1\epsilon$.
The soft push-off potential is now reduced to $0\epsilon$ over a period of $5000\tau_{LJ}$ followed by an energy minimization step.
At this point, all repulsive pairwise interactions (Eq.~\ref{eq:LJ} with $r_{c}=2^{1/6}\sigma$) become active (filler--filler, filler--polymer and polymer--polymer) and the polymer is annealed for $5000\tau_{LJ}$.
This constitutes the end of step 5.

Then, the system is switched to an NPT ensemble at $T=1$ and polymer--polymer interactions are changed to attractive.
The polymer's pressure is initially set to 25 and is then reduced in a stepwise fashion from $25$ to $1$ in LJ units (step 6).
The system is then subjected to an energy minimization.

After this annealing procedure, the polymer chains are cross-linked while the filler particles are frozen in place.
The cross-linking algorithm is then run until 95\% of potential cross-linking bonds are formed, which generally requires around $10,000 \tau_{LJ}$ (step 7).
The system is then isothermally annealed at constant pressure for $5000\tau_{LJ}$.
Then, the pressure is decreased to one, filler--polymer interactions are switched to attractive, and MD simulations are continued in the NPT ensemble for $5000\tau_{LJ}$ to achieve the desired density of the system (step 8), followed by an NVT ensemble simulation for $5000\tau_{LJ}$ (step 9).
Finally, the system is annealed for an additional $5000\tau_{LJ}$ at $P = 0$.


\subsection{Extensional Deformation Simulations in LAMMPS} \label{sec:stretch}


LAMMPS non-equilibrium MD simulations stretch the neat and filled elastomers in this study to 200\% strain.
The elastomer melt is stretched at a constant uniaxial strain rate of $5\times 10^{-5} \tau_{LJ}^{-1}$, which is a rate that is small enough to minimize contributions from high-frequency relaxation modes while being fast enough to allow reasonable wall-clock times for stretching our elastomers to 200\%.~\cite{Smith2017}
This strain rate is 1-2 orders of magnitude lower than rates used in many other studies of polymer deformation.~\cite{Stevens2001, Tsige2004, Hagita2016}
During the deformation in the uniaxial direction, the normal box dimensions are controlled via a zero-pressure anisotropic Nos\'e--Hoover barostat.
Additionally, a constant temperature $T=1$ in LJ units is maintained via a Nos\'e--Hoover thermostat, with a damping parameter of $2\tau_{LJ}$.

To quantify the response to extensional deformation, the resulting stress is stored and analyzed, as in Eq.~\ref{eq:cumm_stress},
\begin{equation} \label{eq:cumm_stress}
    -\sigma^{\alpha \beta} = \frac{1}{V} \left[ \sum_{i=1}^{N} m v^{\alpha}_{i} v^{\beta}_{i} + \sum_{i=1}^{N} W^{\alpha \beta}_{i} \right],
\end{equation}
where $V$ is the initial volume (rendering all stresses in this study as engineering stresses), $N$ is the number of beads, $i$ is the bead number, $\alpha$ and $\beta$ are replaced with Cartesian coordinates $x$, $y$, or $z$, $m = 1$ is the mass, $v$ is velocity, and $W$ is the virial contribution due to intra- and inter-molecular interactions.
This stress tensor is symmetric with six components: $\sigma_{xx}$, $\sigma_{yy}$, $\sigma_{zz}$, $\sigma_{xy}$, $\sigma_{xz}$, and $\sigma_{yz}$.

Eq.~\ref{eq:cumm_stress} can be rewritten to compute the contribution from each bead within the simulation box through $S^{\alpha \beta}$,
\begin{equation} \label{eq:cumm_stress_1}
  \sigma^{\alpha \beta} = \frac{1}{V} \sum_{i=1}^{N} S^{\alpha \beta},
\end{equation}
which results in the relationship,
\begin{equation} \label{eq:bead_stress}
  S_{i}^{\alpha \beta} = -m v^{\alpha}_{i} v^{\beta}_{i} - W^{\alpha \beta}_{i},
\end{equation}
for each bead $i$.
In Equation~\ref{eq:bead_stress}, the first term quantifies the contribution from kinetic energy, while the second incorporates contributions from the potential energy of a bead.

In this work, extensional strain $\varepsilon_{x} = \frac{\left(L_{x}-L_{x,0}\right)}{L_{x,0}}$ is imposed in the $x$ direction, and therefore the stress response is quantified via $\sigma_{xx}$.
Simultaneously, the resulting normal stress is quantified via $P_{\mathrm{norm}} = -\left( \sigma_{yy} + \sigma_{zz} \right)/2$.
To improve the signal-to-noise ratio, which is significant in stress calculations~\cite{Todd2017} especially on a per-bead basis, a time-averaged measurement is stored.
The time-averaged measure of stress takes values every MD step (i.e., every $0.001 \tau_{LJ}$), and then averages them over $2\times10^6$ MD steps or 2000 $\tau_{LJ}$.
With these parameters ($\varepsilon_{x} \in [0, 200\%]$, $\dot{\varepsilon_{x}} = 5 \times 10^{-5} / \tau_{LJ}$ and $\delta t = 10^{-3} \tau_{LJ}$), the simulation takes $\varepsilon_{x}^{\mathrm{Final}}/(\dot{\varepsilon_{x}} \delta t) = 4\times10^{7}$ MD steps.

For all the data points in the main text, we use 15 independent deformation replicates.
Of those 15, three independent filled elastomer systems are prepared according to the protocol previously described.
Then, each of these three systems is thermally forked five times by introducing random velocities to all beads, while maintaining a unit temperature, and then equilibrated for at least three nanoseconds.
This protocol produces 15 thermally forked configurations and ensures that our results are not merely due to stochastic events, which are common in spatially heterogeneous systems.


\subsection{Species Stresses from Deformation Simulations} \label{sec:species}


To capture the intricacies of the stress response, we resolve the measurement on a per species basis.
To do so, a sum over only beads of a certain type (polymer or filler) are computed and then divided by their respective \emph{initial} volume, according to Equation~\ref{eq:species_stress},
\begin{equation} \label{eq:species_stress}
    \sigma^{\alpha \beta}_{I} = \frac{1}{V_{I}} \sum_{i=1}^{N_{I}} S_{i}^{\alpha \beta}
\end{equation}
where $I$ can be polymer or filler, $V_{I}$ is the initial volume of beads $I$, and $N_{I}$ is their number.
$V_{I}$ is computed at the beginning of the simulation using a Voronoi tessellation, and then a sum over the groups' Voronoi volumes.

To facilitate comparisons with the total stress response, as in Figure~\ref{fig:stress_spatial}, we decompose the stress response from Equation~\ref{eq:cumm_stress} as shown in Equation~\ref{eq:species_stress_phi},
\begin{equation} \label{eq:species_stress_phi}
  \begin{split}
    \sigma^{\alpha \beta} & = \frac{1}{V} \sum_{i=1}^{N} S_{i}^{\alpha \beta} = \frac{1}{V} \left[
	  \sum_{i=1}^{N_{P}} S_{i}^{\alpha \beta} + \sum_{i=1}^{N_{F}} S_{i}^{\alpha \beta} \right]  \\
	                  & = \frac{1}{V} \left[
	  \sigma^{\alpha \beta}_{P} V_{P} + \sigma^{\alpha \beta}_{F} V_{F} \right] = \sigma^{\alpha \beta}_{P} \phi_{P} + \sigma^{\alpha \beta}_{F} \phi_{F}
  \end{split}
\end{equation}
where $\phi_{I} = V_{I}/V$ is the volume fraction of the polymer or the filler.
In other words, plotted measures in Figure~\ref{fig:stress_spatial} are all volume fraction $\times$ stress ($\phi^{\mathrm{Filled}} = \phi^{\mathrm{Neat}} = 1$).


\subsection{Poisson Ratio Calculations} \label{sec:poisson}


In these simulations, extensional deformation is applied at a constant engineering rate while maintaining a zero-pressure boundary condition.
This means that $L_{x}$, the box length in the $x$ dimension, is directly varied and $L_{y} = L_{z}$ respond according to the collective Poisson ratio of the matrix, $\nu$, which is given by Equation~\ref{eq:poisson}.
\begin{equation} \label{eq:poisson}
    \nu \equiv - \frac{d \left( \log L_{y} \right)}{d \left( \log L_{x} \right)}
\end{equation}
To compute this, $L_{y}$ values are stored and averaged in 20 equidistant bins (throughout the range $\varepsilon_{x} \in [0, 200\%]$).
Then, for each replicate, finite difference is used to compute $\nu$ via Equation~\ref{eq:poisson}.
Finally, $\nu$ in Figure~\ref{fig:normal_stress}a is the average of $\nu$ from all 15 replicates for the neat and the filled elastomer matrix individually.


\subsection{Stress Response Component Decomposition} \label{sec:component}


To explore reinforcement further, we plot the stress difference $\Delta \sigma = \sigma - \sigma_{0}$ in Figure~\ref{fig:ext_stress}b, where $\sigma$ is the extensional stress of the filled matrix, and $\sigma_{0}$ is the extensional stress of the neat matrix.
In addition to $\Delta \sigma$, Figure~\ref{fig:ext_stress}b presents two other terms, which sum up to the stress difference.
This can be seen via Equation~\ref{eq:stress_comp_methods}, which incorporates the final form of Equation~\ref{eq:species_stress_phi}.
\begin{equation} \label{eq:stress_comp_methods}
  \begin{split}
    \Delta \bm{\sigma} & = \bm{\sigma} - \bm{\sigma}_{0} = \bm{\sigma}_{P} (1-\phi_{F}) + \bm{\sigma}_{F} \phi_{F} - \bm{\sigma}_{0} \\
                       & = \bm{\sigma}_{P} - \bm{\sigma}_{0}\ \ \ \ \ \ + \ \ \ \ \ \ \phi_{F} \left( \bm{\sigma}_{F} - \bm{\sigma}_{P} \right)
  \end{split}
\end{equation}
In Equation~\ref{eq:stress_comp_methods}, the first term $\sigma_{P} - \sigma_{0}$ quantifies the difference in the stress response between the polymer in the filled matrix and the polymer in the neat matrix, i.e., the filler-modified polymer stress contribution, whereas the second term $\phi_{F} \left( \sigma_{F} - \sigma_{P} \right)$ quantifies the direct excess filler stress contribution.


\subsection{Stress Response Spatial Decomposition} \label{sec:spatial}


\begin{figure*}[tbp]
  \includegraphics[]{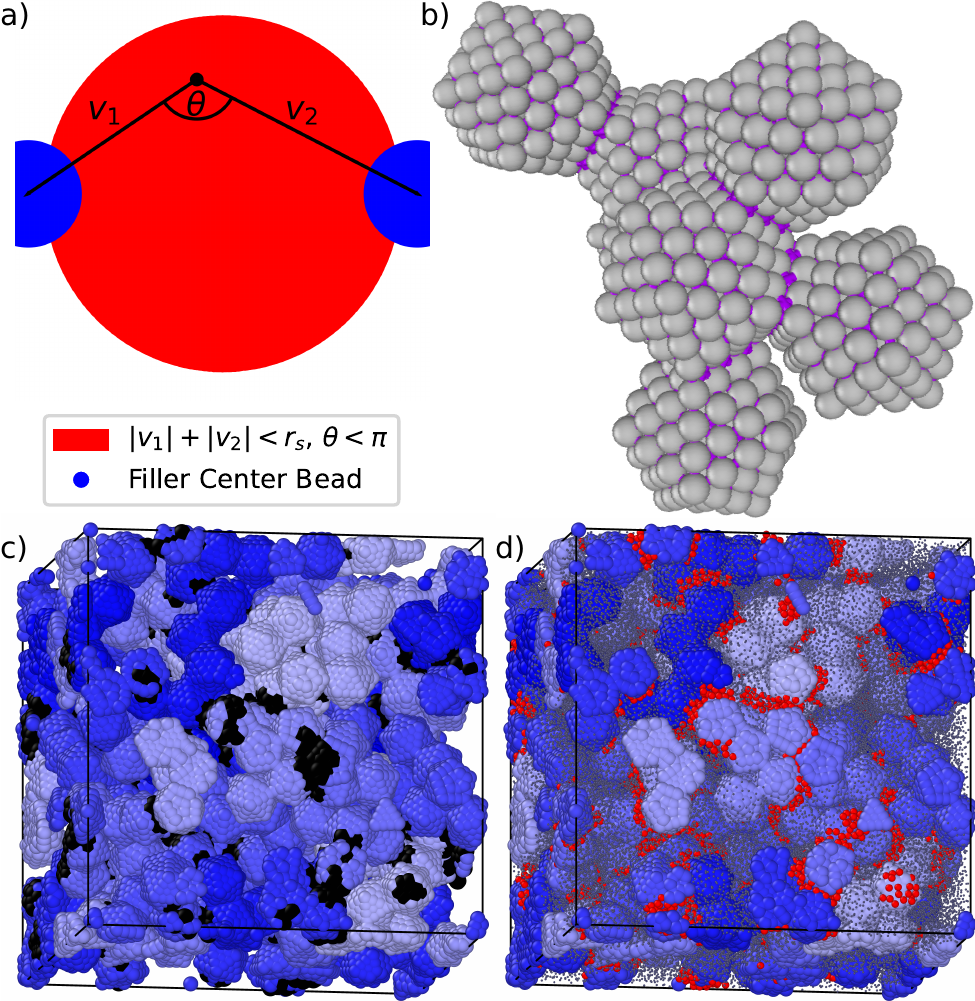}
  \caption{a) Two dimensional illustration of how ``Filler Near-Contact'' and ``Interfiller Polymer'' in Figure~\ref{fig:stress_spatial} are identified.
b) An example of a filler cluster with seven primary icosahedral particles. Filler beads are shown in grey and bonds are in purple.
c) A snapshot showing only filler beads. Beads with the same shade of blue belong to the same cluster and beads labeled black belong to the ``Filler Near-Contact'' region.
d) The same as c) but with polymer beads visible in grey and ``Interfiller Polymer'' region in red.}
  \label{fig:particles_between_scheme}
\end{figure*}

Figure~\ref{fig:stress_spatial} shows the results of a spatial decomposition algorithm that divides all beads into ``interfiller'' and bulk segments.
Specifically, these interfiller beads are sandwiched between two filler particles.
To identify such beads, we developed a simple algorithm that classifies beads based on proximity to filler particles.

The algorithm is illustrated in Fig.~\ref{fig:particles_between_scheme}a.
A bead is classified as ``Interfiller'' if its difference vectors with two filler \emph{center} beads, $v_{1}$ and $v_{2}$, add up to less than some cutoff radius, $r_{s}$, and form an angle, $\theta$, less than $\pi$.
The former condition sets the distance considered close enough, while the latter ensures that the bead exist between the centers and not, for example, by the sides.
This spatial decomposition is employed to distinguish between the validity of the various hypothesized origins of reinforcement.

First, ``Filler Near-Contact'' beads are identified as outer shells of the icosahedral particle that exist within proximity of another particle's center.
The results in the main text used $r_{s} = 15/8\sigma$, which we determined by trial and error to be an appropriate distance between two nearly contacting fillers.
To ensure that permanently contacting beads due to intra-cluster bonding are excluded, the algorithm only compares beads between two filler centers from different filler clusters.
The results from a single snapshot are shown in Figure~\ref{fig:particles_between_scheme}c where ``Filler Near-Contact'' regions are colored in black.

Finally, ``Interfiller Polymer'' beads are identified using the same algorithm.
``Interfiller Polymer'' beads are shown in red in Figure~\ref{fig:particles_between_scheme}d.
Here, $r_{s}=17/8\sigma$ distinguished polymer beads between two neighboring filler particles.
These results are used in the main text to distinguish between the relative roles of the glassy bridges reinforcement model~\cite{Dannenberg1986, Merabia2008, Papon2012, Starr2016, Tauban2017} and the direct contact model.~\cite{Zhu2005, Witten1993}

To obtain the necessary resolution for these ensemble averages of atomic stress, we saved a snapshot every 0.02\% extensional strain (a total of $10^4$ snapshots) from 15 distinct extensional deformation simulations.
Regions were dynamically identified for each snapshot and the atomic stresses of resident beads were averaged to give the responses in Figure~\ref{fig:stress_spatial}.

\section{Results}

\begin{figure}[htb]
  \centering
  \includegraphics[width=0.82\columnwidth]{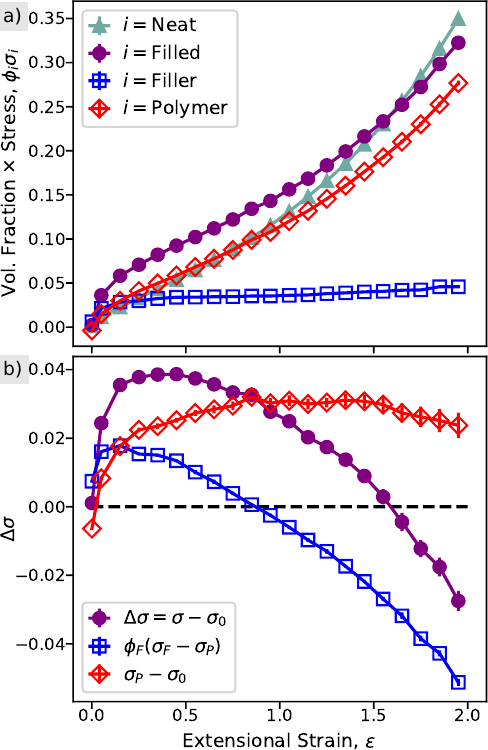}
  \caption{Engineering extensional stress--strain curves from extensional deformation of filled elastomers showing a) the product of volume fraction $\phi$ and extensional stress, $\sigma$ ($\phi_{\mathrm{Filled}}=1$), and b) the change in stress between filled and neat elastomers $\Delta \sigma$, as well as its two decomposed components (as described in the text). All scatter points include stress standard error bars calculated from 15 distinct deformation simulations.}
  \label{fig:ext_stress}
\end{figure}

Figure~\ref{fig:ext_stress}a reports on three regimes of modified nanocomposite stress response relative to the neat polymer.
As previously reported~\cite{Smith2017, Smith2019}, at low strains the nanocomposite exhibits an appreciable enhancement in stress response and Young's modulus.
In the vicinity of strain where the Payne effect is expected (\textit{ca.} 10\%--20\%), the system exhibits strain softening, with the stress response afterwards exhibiting a nearly vertical shift relative to the pure system.
Finally, the system exhibits a surprising crossover in the vicinity of 150\% strain, beyond which the neat elastomer exhibits a \emph{higher} stress than the nanocomposite.
This final crossover is characteristic of the low-to-intermediate loading regime for these simulations that lack strong hydrodynamic effects, whereas at higher loadings, more characteristic of highly reinforced elastomers, this crossover is lost.
However, as we discuss below, its presence at intermediate loading is a signature of the physics of nonlinear reinforcement.

To understand this behavior, we decompose the stress response of the nanocomposite into component stresses $\bm{\sigma}_{P}$ and $\bm{\sigma}_{F}$ for the polymer and nanofiller respectively. A volume-weighted average of these contributions then yields the total stress (see Methods Section for details):
\begin{equation} \label{eq:stress_species}
\bm{\sigma} = \phi_{P}\bm{\sigma}_{P} + \phi_{F}\bm{\sigma}_{F},
\end{equation}
where $\phi_{P}$ and $\phi_{F}$ are the polymer and filler volume fractions, respectively. As can be seen in Figure~\ref{fig:ext_stress}a, this decomposition reveals a qualitative change in the nature of the direct contribution from the filler beginning around \textit{ca.} 20\% strain. At lower strain, the filler network contributes to the elastic response of the system at a level comparable to the polymer response.
At higher strains, the filler stress contribution plateaus, which is characteristic of a purely dissipative response.
This crossover in the nature of nanoparticulate component stress is consistent with expectations for a yield of the filler network corresponding to the Payne effect.

We can better discern the precise mechanism of reinforcement if we consider direct polymer and nanoparticulate contributions to the \emph{difference} between the neat and nanocomposite elastomers (see Figure~\ref{fig:ext_stress}b) as in Equation~\ref{eq:stress_comp} (see Methods Section for details).
\begin{equation} \label{eq:stress_comp}
\Delta \bm{\sigma} \equiv \bm{\sigma} - \bm{\sigma}_{0} = \phi_{F}\left( \bm{\sigma}_{F} - \bm{\sigma}_{P} \right) + \left(\bm{\sigma}_{P} - \bm{\sigma}_{0}\right)
\end{equation}
The resulting decomposition, shown in Figure~\ref{fig:ext_stress}b, indicates a two-part mechanism of reinforcement involving both a direct filler contribution to composite stress and an indirect (mediated by polymer) filler contribution to the composite stress.
The direct filler contribution, quantified by $\phi_{F}\left( \bm{\sigma}_{F} - \bm{\sigma}_{P} \right)$, reports on the stress contribution associated with replacing $\phi_{F}$ volume fraction of polymer with NPs.
This term contributes to an enhanced composite modulus at low strains before the Payne effect.
Beyond the Payne effect, however, the filler contribution reverses, such that the filler most directly \emph{reduces} the composite stress.
The nonmonotonicity of the filler's direct contribution in Figure~\ref{fig:ext_stress}b indicates that, for high strains, this dissipative contribution is insufficient to offset the loss of elastic contribution from the polymer that the NPs replace.
This is a natural outcome of replacing part of an elastic medium with a dissipative medium, since elastic stresses innately grow with strain whereas dissipative stresses do not.
In fact, this is what drives the crossover to net softening around 150\% strain.

In parallel to its direct contribution, the presence of NPs also enhances the stress response within the polymer itself.
The filler-modified polymer response, quantified by $\bm{\sigma}_{P} - \bm{\sigma}_{0}$, reports on the indirect contribution of NPs in modifying the polymer mechanical response.
As can be seen in Figure~\ref{fig:ext_stress}b, this term contributes a large enhancement in the low strain regime, and \emph{remains elevated at a near-plateau far after the Payne effect}.
This is evidently the dominant mechanism driving stress enhancement at high strain.
What gives rise to this second mechanism that keeps the polymer's response elevated after the filler matrix has dejammed in the elongational direction?

\begin{figure}[htb]
  \centering
  \includegraphics[width=0.82\columnwidth]{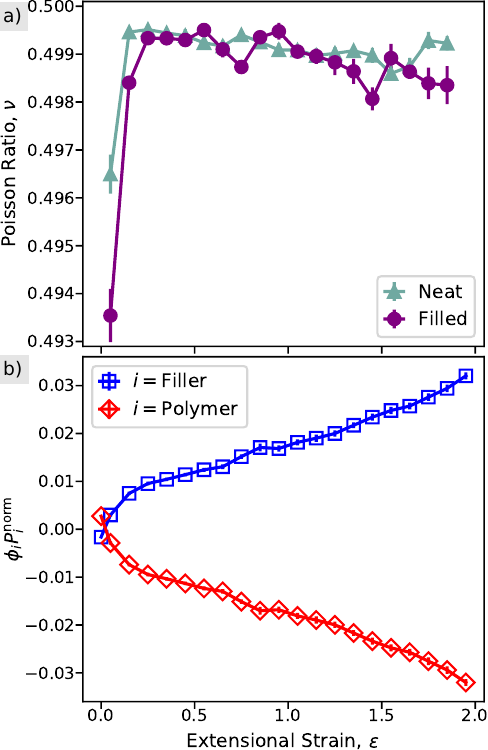}
  \caption{a) Poisson ratio, $\nu$, from extensional deformation simulations of neat and filled elastomers. b) Product of species (filler and polymer) volume fraction, $\phi$, and pressure in the normal direction, $P^{\mathrm{norm}}$, versus extensional strain, $\varepsilon$, in the filled composite. All scatter points include standard error bars calculated from 15 distinct deformation simulations.}
  \label{fig:normal_stress}
\end{figure}

As discussed in our prior work~\cite{Smith2019} and indicated by Figure~\ref{fig:normal_stress}a, the initial large enhancement in $\sigma_{P}$ is caused by a downward deviation in the composite Poisson's ratio $\nu$.
This deviation is induced by the filler network, which prefers to increase its volume upon stretching.
In turn, this drives a contribution to the polymer's elongational stress response from its bulk modulus to resist volume expansion.
As shown by this figure, after yield, $\nu$ recovers to near (but not quite equal to) the neat elastomer Poisson ratio $\nu_{0} \sim 0.5$.
However, this recovery is insufficient to restore the polymer's preferred equilibrium volume.
Indeed, recovery to a transient Poisson ratio of 0.5 at these higher strains only maintains the nonequilibrium (increased) volume of the system attained during the early phases of strain;
restoration of the equilibrium volume would require volume contraction (corresponding to deformation at a Poisson ratio greater than 0.5).
This evidently does not occur due to ongoing resistance to normal contraction by the nanoparticle network.

This deviation from the polymer's desired volume drives a large internal normal stress balance $P^{\mathrm{norm}}$, as shown in Figure~\ref{fig:normal_stress}b, wherein the $P^{\mathrm{norm}} = 0$ condition emerges from a compensation between $P^{\mathrm{norm}}_{F} > 0$ and $P^{\mathrm{norm}}_{P} < 0$.
The rate of growth of this internal normal stress competition weakens after the Payne effect, but a large internal stress balance persists.
Restoring polymer equilibrium volume would require deformation at $\nu > \nu_{0}$ in order to `make up' for the accumulated positive deviation from the equilibrium polymer volume.
As per Figure~\ref{fig:normal_stress}b, this evidently does not occur due to continued outward stress imposed by the fillers.
As a consequence, additional work is required to continue elongating the sample against the polymer's effort to isotropically contract, such that the consequences of the low-strain Poisson ratio mismatch physics continue to drive reinforcement well into the nonlinear regime.

In essence, these findings point to a feedback loop between polymer incompressibility and nanofiller granular deformation.
The filler induces a growth in polymer volume that invokes a contribution from the polymer's bulk modulus, reinforcing the response at low strains.
At high strains, this effect leads to lateral compression of the filler, maintaining a near-jammed NP state and enabling a post-yield plastic elongational response that would be absent in the neat polymer.

\begin{figure*}[htb!]
  \centering
  \includegraphics[]{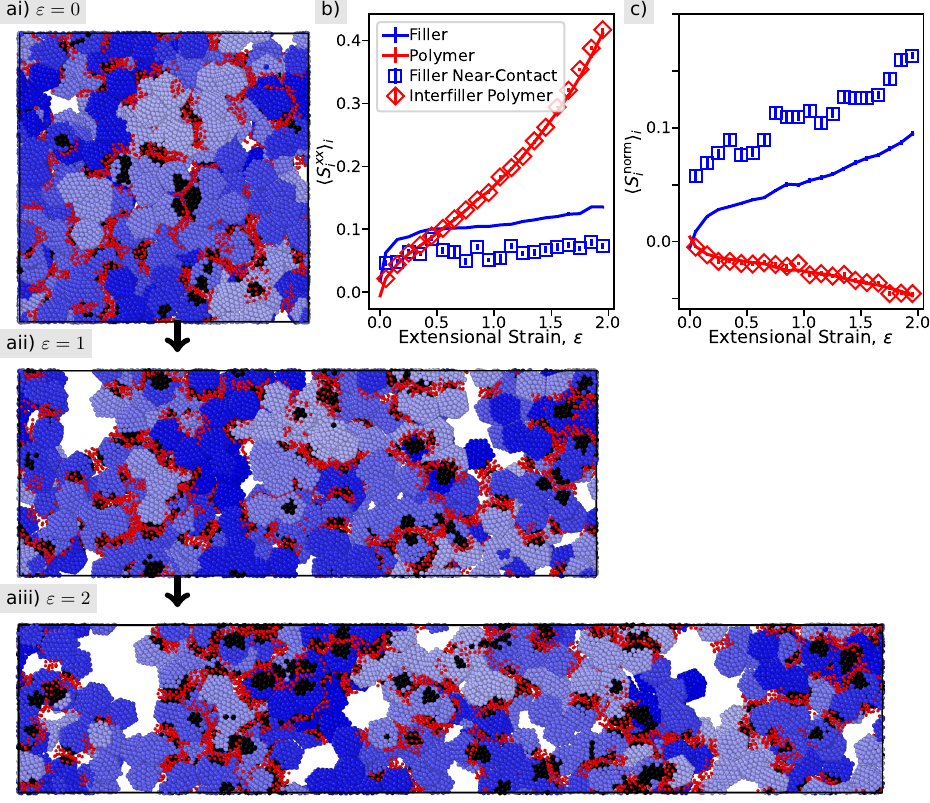}
  \caption{a) Snapshots from stretching MD simulations at $\varepsilon=0.0$ (top, ai), $\varepsilon=1.0$ (middle, aii), and $\varepsilon=2.0$ (bottom, aiii) created with OVITO visualization software.~\cite{Stukowski2009} Filler particle clusters are displayed in different shades of blue. To illustrate our stress spatial decomposition algorithm, filler beads near particle contacts (``Filler Near-Contact'') are shown in black and polymer beads between two filler particles (``Interfiller Polymer'') are shown in red. b) Spatially-resolved extensional stress response from ``Interfiller Polymer'' and ``Filler Near-Contact'' regions. c) Same as b, but with normal pressure. All scatter points include standard error bars calculated from 15 distinct deformation simulations.}
  \label{fig:stress_spatial}
\end{figure*}

We can resolve the mechanism for this feedback loop even more clearly by computing a per-bead contribution to stress.
We then average this quantity over only those filler beads that are near contact points between distinct filler particles, and we compare this to an average over all filler beads throughout the system.
We similarly perform this analysis for polymer beads that are located between closely-spaced filler particles and compare to an average over all polymer beads.
These local bead determinations are reperformed at each time step so that they are updated as the sample deforms.
The result of this scheme is depicted in the snapshots shown in Figure~\ref{fig:stress_spatial}a, which shows how the near-contact and interfiller regions evolve with deformation.

As can be seen in Figure~\ref{fig:stress_spatial}b and~\ref{fig:stress_spatial}c, the stress response of polymer beads located between NPs is essentially indistinguishable from the mean polymer response of the system.
This confirms that the enhancement in polymer-domain stress contributions is essentially homogeneous and is not localized between the particles; this is consistent with the bulk-modulus mechanism described above but would be inconsistent with any polymer bridging or strain localization mechanism.
Critically, we find that this mechanism can drive stress enhancement in the absence of effects such as glassy polymer bridging and hydrodynamic reinforcement, indicating that those effects may abet reinforcement but are not a necessary ingredient.

In contrast, filler near-contact regions exhibit distinct stress behavior from the filler average: filler stress is \emph{reduced} near contact points in the elongational direction, but \emph{enhanced} in the normal direction (c.f., Figure~\ref{fig:stress_spatial}b and~\ref{fig:stress_spatial}c).
The reduction in filler near-contact point elongational stress explains why the direct contribution from NPs at high strain is primarily dissipative rather than elastic: the NPs are simply not strongly cohesive and do not directly elastically resist high-strain elongation, as in the analogy of concrete.
Conversely, the enhanced normal stress at near-contact points emphasizes that the filler network continues to bear load in the normal direction under high strain, maintaining the polymer's non-equilibrium volume and enabling the Poisson-ratio-mismatch physics described above to continue abetting elongational reinforcement to high strains.

\section{Discussion and Conclusions}

For nearly a century, the precise origin and mechanisms of nanoparticulate reinforcement of elastomeric mechanical response have been a major open question in polymer and nanomaterials science.
These results demonstrate that nonlinear reinforcement emerges from a complex interplay of polymer physics and granular jamming yield physics.
At both low and high strains, we find that elongational reinforcement is dominated by a competition between the system volume preferred by the polymer matrix and the nanoparticulate filler.
Within this competition, the filler network compels the polymer matrix to sustain positive deviations from it equilibrium volume, invoking a contribution from the (relatively very large) matrix bulk modulus to the composite Young's modulus.
In turn, the tendency of the polymer to contract in the normal direction maintains a near-jammed nanofiller state well beyond its expected yield point.
These physics lead to a large internal normal stress balance wherein the polymer matrix is effectively at large negative pressure, with this negative pressure balanced out by a positive filler pressure that we observe to be mediated via direct filler contacts.
We expect that, at even higher strains than those probed here, this will play a central role in mediating and modifying void nucleation and ultimate failure.
Our evidence suggests that this internal normal stress balance is thus a defining and controlling feature of elastomeric reinforcement by nanoparticles.

These findings shed new light on the debate over the dominant mechanistic origin of reinforcement.
Our results indicate that `glassy bridge' or bridging chain effects are not necessary to yield strong reinforcement.
Instead, our results support the core physical proposition of the WRC model:~\cite{Witten1993} reinforcement of elastomeric nanocomposites at high strain is mediated by ongoing lateral compression of a network of directly contacting nanoparticles. Future simulations should build on this foundation by comparing to the quantitative predictions of the WRC model. This will likely require simulations spanning a range of filler structures and loadings, so as to enable physically meaningful exploration of a scaling exponent in the WRC theory that is interpreted in terms of filler network structure. Such work could also provide more quantitative insight into the extent to which reinforcement at intermediate to high strains is coupled to low-strain (i.e. Young's modulus) reinforcement. Our present findings suggest that at a mechanistic level reinforcement in both of these regimes is dominated by particle-network-driven deviations from equilibrium polymer volume. This may suggest a relatively strong correlation between the quantitative degree of reinforcement before and after the Payne effect; a test of this proposition will require a much more structurally diverse data set.

At the same time these findings support the qualitative physical scenario of the WRC model, results identify multiple points on which an updated theoretical understanding should build on this foundation.
First, an updated theory must account for the deviations from incompressibility that we find play a central mediating role in reinforcement physics.
Second, we find that the stress response of the filler in the elongational direction becomes entirely dissipative, rather than elastic as suggested by the WRC scenario, after the Payne effect (Figure~\ref{fig:ext_stress}).
While this may change in particles with stronger explicit cohesive interactions, our results indicate that no direct elongational elasticity of the nanoparticles themselves is necessary for reinforcement and is likely a second-order effect.
Updates to a theoretical description of this phenomenon thus should not intrinsically require direct elongational elasticity of the nanoparticles.
Moreover, these findings suggest that the structural and flocculation state of the filler, and their roles in mediating percolation of direct filler--filler contacts, are likely to be the most central determinant of reinforcement.
Future work developing a more predictive understanding of this relationship would thus be of high value in the development of nanocomposites with targeted mechanical properties.
We furthermore expect that introduction of so-called ``glassy bridge'' or ``bound rubber'' effects at lower temperatures and stronger interactions may augment the mechanism reproted here, perhaps by abetting the tendency of the particle network to bear compressive load in the normal direction;
we expect to explore these potential effects in future work.

Finally, this new understanding of elastomeric nanocomposite reinforcement raises essential questions regarding how this mechanism evolves at even higher strains approaching failure.
Failure of nanofilled elastomers can involve nucleation of voids that play a central role in cracking.
How do these mechanisms alter void nucleation?
To what extent is void formation nucleated at the particulate surface, and to what extent do particle--polymer attractive interactions play an emergent role in this higher strain regime?
Future work answering these questions and extending this understanding to material failure could provide additional predictive insights into the design of ultra-tough elastomeric nanocomposites.

\begin{acknowledgement}
This material is based upon work supported by the U.S. Department of Energy, Office of Science, Office of Basic Energy Sciences, under Award Number DE-SC0022329.
\end{acknowledgement}

\clearpage
\bibliography{refs}

\clearpage

\end{document}